\documentclass[times,authoryear,onecolumn,final]{elsarticle}

\usepackage{jasr}
\usepackage{framed,multirow}

\usepackage[table,xcdraw]{xcolor}

\usepackage{colortbl}
\usepackage{array}

\usepackage{amssymb}
\usepackage{latexsym}
\usepackage{subcaption}
\usepackage{soul}

\usepackage[switch]{lineno}

\usepackage{url}
\usepackage{xcolor}
\definecolor{newcolor}{rgb}{.8,.349,.1}
\DeclareUnicodeCharacter{2009}{\,}

\usepackage[citebordercolor=white]{hyperref}

\journal{Advances in Space Research}

\begin{document}

\verso{Ayisha M Ashruf \textit{etal}}

\begin{frontmatter}

\title{Loss of 12 Starlink Satellites Due to Intense Space Weather Activity Surrounding the Extreme Geomagnetic Storm of 10 May 2024}

\author{Ayisha \snm{M Ashruf}\textsuperscript{a,b}\corref{cor1}}
\cortext[cor1]{Corresponding author: }
\ead{ayisha@vssc.gov.in, ayishamashruf@gmail.com}
\author{Ankush \snm{Bhaskar}\textsuperscript{a}}
\author{C. \snm{Vineeth}\textsuperscript{a}}
\author{Tarun \snm{Kumar Pant}\textsuperscript{a}}
\author{Ashna \snm{V M}\textsuperscript{c}}

\affiliation[1]{organization={Space Physics Laboratory},
	addressline={Vikram Sarabhai Space Centre},
	city={Thiruvananthapuram, Kerala},
	postcode={695022},
	country={India}}
\affiliation[2]{organization={Indian Institute of Space Science and Technology},
	addressline={Valiamala},
	city={Thiruvananthapuram, Kerala},
	postcode={695547},
	country={India}}
\affiliation[3]{organization={Department of Physics, Providence Women's College, University of Calicut},
	addressline={Malaparamba},
	city={Kerala},
	postcode={},
	country={India}}


\begin{abstract}
	
Low-Earth orbit satellites are highly sensitive to space weather, which can significantly alter their trajectories through increased atmospheric drag. In this study, we analyze the orbital decay of 12 Starlink satellites between 16 April and 15 May 2024 using Two-Line Element (TLE) data. Our analysis reveals two distinct phases of decay: one characterized by a pronounced increase in orbital decay following the extreme geomagnetic storm on 10 May 2024, and an unexpected preconditioning phase — with enhanced decay around 25 April 2024 — observed in satellites above 320 km. We attribute the pre-storm period decay to an increase in solar Extreme Ultraviolet (EUV) output, coinciding with a series of M class flares on the Sun and an associated thermospheric density enhancement leading to increased atmospheric drag. Furthermore, all of these Starlink satellites exhibited sharp decay during the early recovery phase of the geomagnetic storm. This study highlights the complex role of preconditioning, driven by enhanced EUV flux and extreme geomagnetic activity, in shaping the orbital dynamics of Low-Earth Orbit (LEO) satellites, and underscores the need for a better understanding of their impact, especially as more satellites are launched into LEO.
	
\end{abstract}


\end{frontmatter}


\section{Introduction}

Satellites play a critical role in global communications, Earth observation, navigation, and scientific research. However, they are vulnerable to space weather disturbances, especially those at LEO altitudes. Understanding the impacts of these disturbances is essential for ensuring the longevity of satellites. Atmospheric drag is the primary force driving the orbital decay of satellites, and it depends on several factors: the local atmospheric density, the satellite's cross-sectional area, its relative velocity with respect to the atmosphere, and the drag coefficient. The drag coefficient quantifies the interaction between the surface of the spacecraft and the impinging atmospheric molecules within the free-molecular flow regime \citep{DOORNBOS2006}. Among these factors, atmospheric density is the most significant variable influencing drag, and it is largely driven by solar activity.

Eruptive phenomena on the Sun, such as solar flares and coronal mass ejections (CMEs), inject energy into Earth’s magnetosphere and ionosphere, leading to thermospheric density and temperature enhancements. When charged particles from CMEs reach Earth, they interact with the magnetosphere, generating increased currents in the magnetosphere-ionosphere system and causing Joule heating, particle precipitation, and thermospheric expansion \citep{Sutton2009}. This can result in a substantial increase in atmospheric density at LEO altitudes, particularly during geomagnetic storms, where the total mass density can rise by more than an order of magnitude \citep{Forbes2005}. For LEO satellites, this increased density leads to greater drag forces, which can alter their trajectories, reduce operational lifespans, and necessitate more frequent orbit maneuvers \citep{zesta2019thermospheric}. Furthermore, \cite{Oliveira_2020} has quantified the possible impact of historic extreme geomagnetic storms (March 1989, October/November 1903, September 1909, and May 1921) on satellite drag, demonstrating the potential for severe orbital decay. Disruptions from space weather events can thus have significant impacts on satellite operations, affecting applications ranging from satellite navigation to weather forecasting.

Solar Cycle 25 reached its peak on 15 October 2024 (\url{svs.gsfc.nasa.gov/14683/}), characterized by frequent geomagnetic disturbances and aurorae visible at lower latitudes \citep{vichare2024low}. On 1 May 2024, Active Region 3664 (AR3664) emerged on the southeastern limb of the Sun's visible disk. It rapidly intensified between 3 and 6 May, producing multiple X-class solar flares and Earth-directed CMEs. This activity triggered one of the most powerful geomagnetic storms in the past two decades, which struck Earth on 10 May 2024 \citep{hayakawa2024,lazzus2024report,Thampi2025}. The storm's effects were severe, causing aurorae to be observed even at mid-latitudes and uplifting the ionosphere to higher altitudes \citep{karan2024gold,thampi2024super,kwak2024observational}. Furthermore, the storm impacted the Earth’s radiation belts, resulting in a strong Forbush decrease in cosmic ray intensity \citep{Pierrard2024,mavromichalaki2024unusual}. The impact of the May 2024 Gannon storm on satellite drag was reported by \cite{parker2024satellite}, noting that it caused significant and unpredictable perturbations in the trajectories of satellites in LEO. They also reported that non-maneuverable debris and defunct satellites experienced accelerated orbital decay, particularly at altitudes between 400 and 700 km, resulting in a temporary `debris sink' that provided a rare environmental benefit by passively reducing space debris.

Based on data from the Space-Track website (\url{https://www.space-track.org}), it was found that 58 orbiting bodies reentered Earth's atmosphere from 8 to 15 May 2024. Among them, 12 were Starlink satellites (Table \ref{tab:Tab1}). Out of the 58 objects, 39 were in high-inclination to near-polar orbits, while the remaining 19, including the Starlink satellites, had inclinations of less than 53$^{\circ}$.  In comparison, only 30 objects were recorded with decay dates between 1 and 7 May 2024. In this study, we investigate the orbital decay of the 12 Starlink satellites during the extreme geomagnetic storm of 10 May 2024. A Starlink satellite has a typical lifespan of around 5 years. Among the 12 Starlink satellites analyzed in this study, one had been in orbit for a year, one for less than two years, eight for less than three years, and two for four years. Hence, it is highly likely that the geomagnetic disturbances associated with the storm caused accelerated reentry of these satellites. Using TLE data, we examined the orbital decays of the satellites before, during, and after the storm. It was seen that the Starlink satellites that were not affected were at higher altitudes, above 500 km, while the 12 satellites included in this study were located at altitudes between 300 and 400 km. This study offers a unique opportunity to assess the impact of an extreme storm on LEO satellites, as no similarly severe storm has affected satellites in the past two decades. The findings provide valuable insights into how pre-storm solar conditions and their effect on thermospheric density can influence satellite drag and orbital decay rates.

\section{Data and Methodology}

As mentioned in the introduction, this study examines the orbital decay of 12 Starlink satellites during the period from 16 April to 15 May 2024 by integrating TLE data, EUV flux measurements, and atmospheric density models.
TLEs summarize the orbital parameters into a maximum of 69 alphanumeric characters. A detailed description of the TLE format can be found in \cite{VALLADO2012}. TLE data from 16 April 2024 to 15 May 2024, for the objects listed in Table \ref{tab:Tab1} were obtained from the Space-Track website. Based on Kepler's third law, the square of the orbital period of an orbiting body is directly proportional to the cube of the semi-major axis of its orbit. The semi-major axis ‘a’ of the orbit of each satellite at a given epoch is obtained from mean motion using the following equation, 

\begin{eqnarray}
	a = \left(\frac{GM}{\left(\frac{2\pi n}{86400}\right)^{2}}\right)^{\frac{1}{3}}
\end{eqnarray}

where $M$ is the mass of the Earth, $G$ is the universal gravitational constant, and $n$ is the mean motion in revolutions per day, as provided in the TLE dataset. 

\begin{table}[htbp]
	\centering
	\caption{Satellite ID, inclination, and slopes across three regions: inc - inclination, m1 - slope in region 1, m2 - slope in region 2, m3 - slope in region 3}
	\begin{tabular}{
			>{\columncolor[HTML]{FFFFFF}}c 
			>{\columncolor[HTML]{FFFFFF}}c 
			>{\columncolor[HTML]{FFFFFF}}c 
			>{\columncolor[HTML]{FFFFFF}}c 
			>{\columncolor[HTML]{FFFFFF}}c }
		\hline
		{\color[HTML]{000000} NORAD ID} & {\color[HTML]{000000} inc($^\circ$)} & {\color[HTML]{000000} m1 (km/day)} & {\color[HTML]{000000} m2 (km/day)} & m3 (km/day) \\ \hline
		57649                           & 42.98                      & -0.73                                   & -8.13                                   & -45.37           \\
		48384                           & 53.04                      & -1.08                                   & -5.00                                   & -56.40           \\
		48599                           & 53.04                      & -0.14                                   & -5.58                                   & -61.47           \\
		48324                           & 53.04                      & -1.47                                   & -5.49                                   & -71.41           \\
		44951                           & 52.99                      & -2.27                                   & -                                       & -49.51           \\
		54001                           & 53.21                      & -0.07                                   & -5.32                                   & -46.82           \\
		48430                           & 53.07                      & -0.14                                   & -5.42                                   & -75.95           \\
		47915                           & 53.02                      & -0.10                                   & -5.53                                   & -58.22           \\
		44928                           & 53.04                      & -2.05                                   & -                                       & -57.44           \\
		48579                           & 53.04                      & 0.42                                    & -5.35                                   & -46.62           \\
		48308                           & 53.03                      & -0.06                                   & -5.71                                   & -49.65           \\
		47741                           & 53.02                      & -0.02                                   & -5.57                                   & -51.16           \\ \hline
	\end{tabular}
	
	\label{tab:Tab1}
\end{table}

Atmospheric temperature, composition, and density are strongly influenced by space weather effects, particularly variations in solar EUV radiation and CMEs \citep{DOORNBOS2006}. EUV flux data is crucial for understanding the changes associated with variations in solar output that can result in the variability of the ionosphere-thermosphere region. GOES EUVS \citep{Eparvier2009} measures EUV and far ultraviolet (FUV) high-spectral-resolution measurements of distinct solar emission lines representative of different layers of the solar atmosphere. The absorption of EUV (10–120 nm) and UV (120–200 nm) radiation accounts for roughly 80\% of the energy entering into the thermosphere. This process raises the temperature of the thermosphere, leading to its upwelling/downwelling as the EUV irradiance increases/decreases with the movement of active regions across the solar disk \citep{Vourlidas2018}. Since EUV radiation is one of the primary energy inputs at LEO altitudes, influencing both the geospace environment and satellites, the EUV flux data from Extreme Ultraviolet Sensor (EUVS) on-board the Extreme Ultraviolet and X-ray Irradiance Sensors (EXIS) on the GOES-16 satellite (\url{https://data.ngdc.noaa.gov/platforms/solar-space-observing-satellites/goes}) between 16 April 2024 and 15 May 2024 is compared with the orbital decays of the 12 satellites.

The interplanetary conditions were monitored using the ACE and Wind satellites, which observe upstream solar wind, while geomagnetic activity is tracked by ground-based magnetometer observatories worldwide. Data on solar wind and interplanetary magnetic fields are obtained from CDAWEB (\url{https://cdaweb.gsfc.nasa.gov/}), and the provisional Dst index and auroral indices are sourced from the World Data Center in Kyoto (https://wdc.kugi.kyoto-u.ac.jp/).

Orbital decay profiles were generated by plotting TLE-derived satellite altitudes over time. These profiles were then compared with the corresponding variations in EUV flux and Dst index to identify two distinct decay phases: a preconditioning phase prior to the geomagnetic storm and a rapid decay phase during the storm’s recovery.

\section{Observations}
\subsection{Solar and Interplanetary Conditions}

\begin{figure}[ht]
	\centering
	\includegraphics[width=0.95\textwidth]{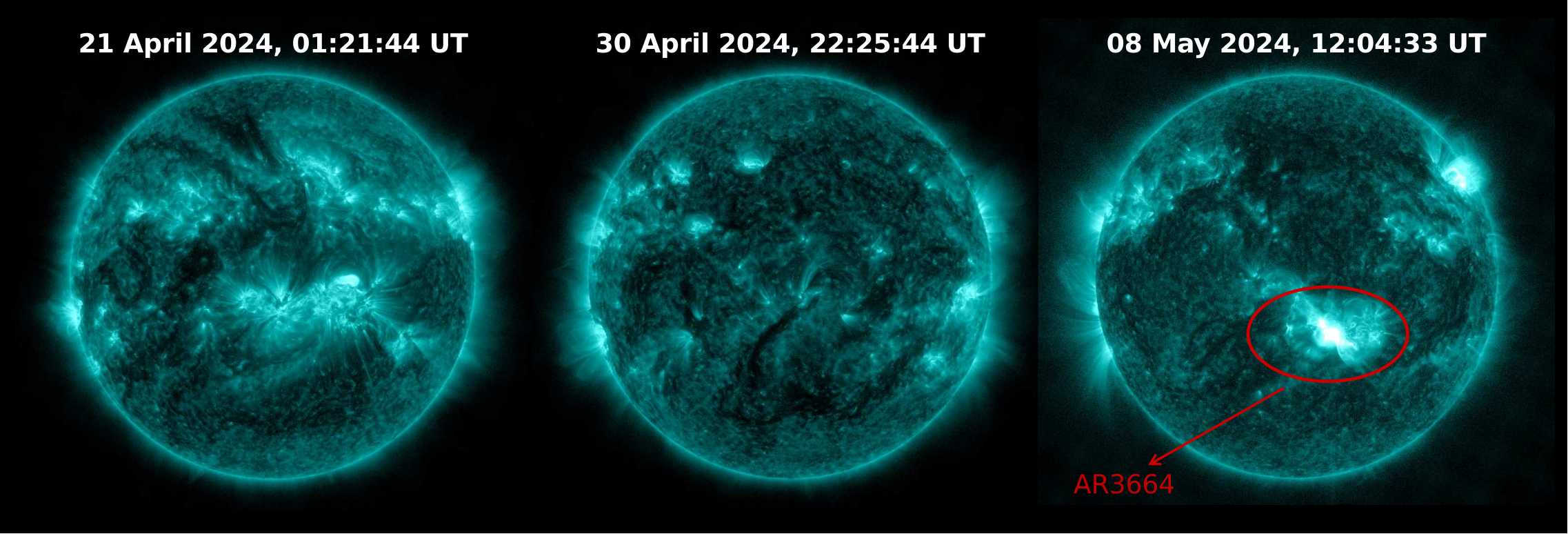}
	\caption{Solar disk UV emission in 131 nm wavelength during April-May 2024, as observed by SDO with the active region AR3664 shown enclosed in a red circle in the rightmost panel.}
	\label{fig:SDODiskImages}
\end{figure}

The increased magnetic activity on the Sun led to the formation of numerous active regions and increased UV and EUV emissions. Figure \ref{fig:SDODiskImages} shows snapshots of the solar disk captured by the SDO in UV emissions at 131 nm. It is evident that the solar disk exhibited multiple bright emission regions around 21 April, with decreased emissions observed around 30 April, and then a subsequent increase in emissions by 8 May 2024. The emergence of active region AR3664 on the Sun is seen in the rightmost image.

\begin{figure}[ht]
	\centering
	\includegraphics[width=0.8\textwidth]{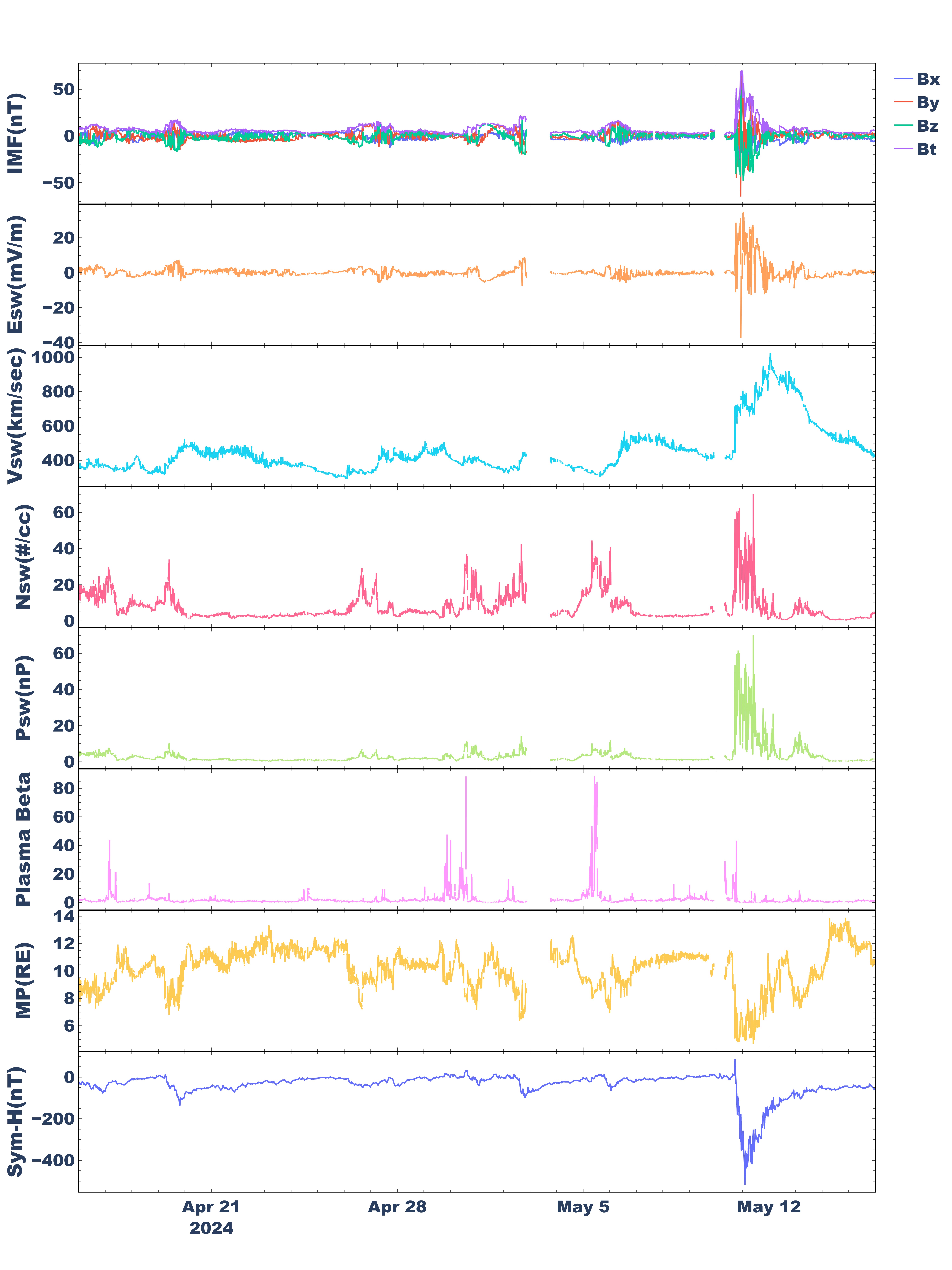}
	\caption{Interplanetary magnetic field parameters and solar wind velocity between 16 April 2024 and 15 May 2024. From top to bottom, Interplanetary Magnetic Field GSM components (Bx, By, Bz) and total field strength (B$_{t}$), solar wind electric field( E$_{sw}$), speed (V$_{sw}$), density (N$_{sw}$), dynamic pressure(P$_{sw}$), subsolar magnetopause location (MP), Symmetric ring current index (Sym-H).}
	\label{fig:IP_long}
\end{figure}

Figure \ref{fig:IP_long} depicts the solar wind, interplanetary magnetic field (IMF), and geomagnetic parameters during the study period. The most pronounced perturbations occurred on 10 May 2024, with IMF strength exceeding 60 nT, solar wind speeds surpassing 800 km/s, density and solar wind dynamic pressures peaking at around 60 cm$^{-3}$ and 60 nPa respectively. The estimated subsolar magnetopause location, determined using the method described in \cite{Shue_1998}, indicates that the magnetopause moved inward to below 6 R$_{E}$. Prior to the storm, minor fluctuations in IMF and solar wind parameters were observed, and the Sym-H index showed no significant decrease, indicating relatively low geomagnetic activity before the onset of the intense storm.

\subsection{ Impact on Starlink Satellite Orbits}

\begin{figure}[ht]
	\centering
	\includegraphics[width=0.85\textwidth]{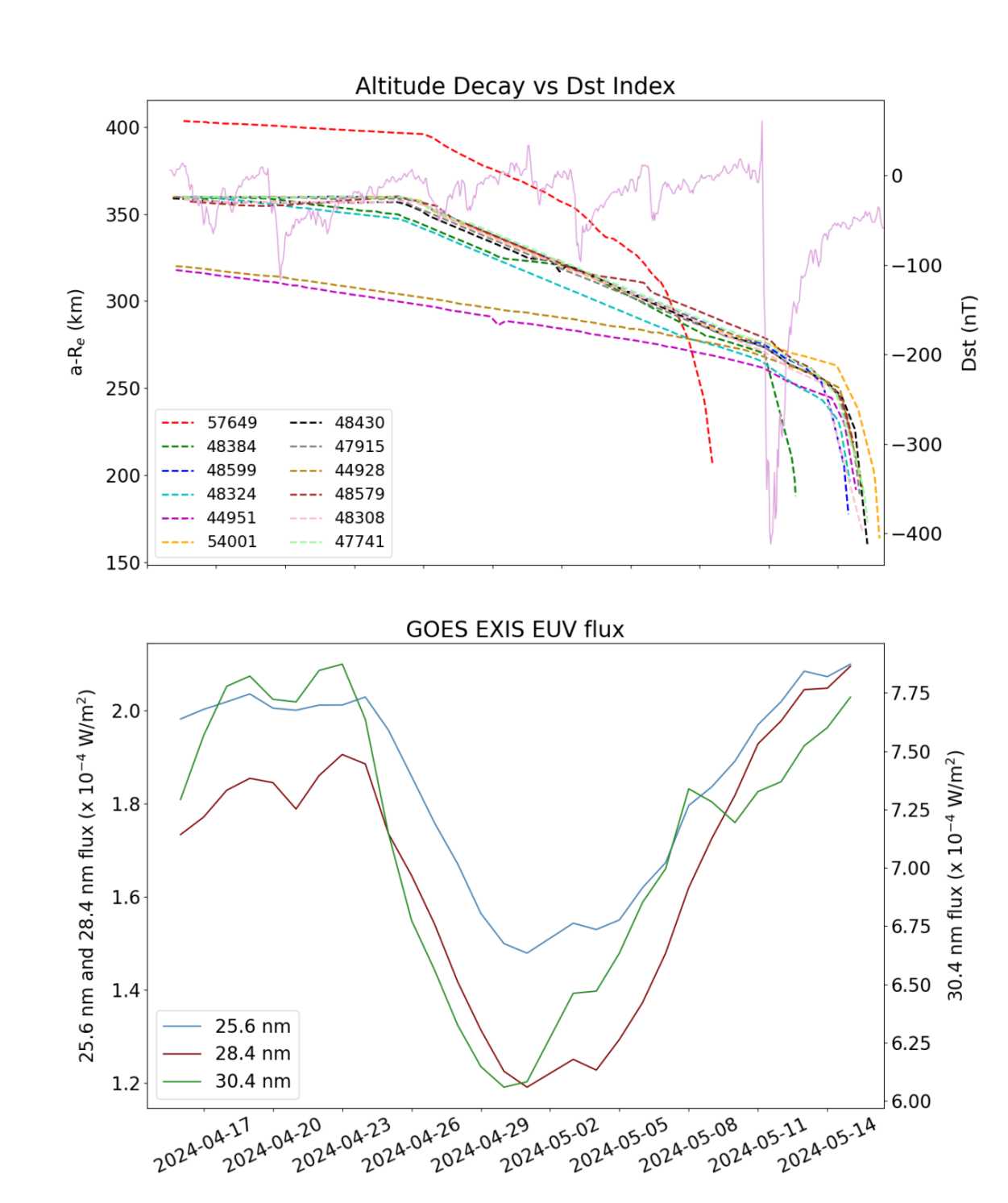}
	\caption{TLE derived altitude decay profiles of the 12 Starlink satellites between 16 April 2024 to 15 May 2024 along with Dst index and EUV flux data from GOES/EXIS Extreme Ultraviolet Sensor (EUVS)}
	\label{fig:DecayProfile}
\end{figure}

\begin{figure}[ht]
	\centering
	\includegraphics[width=0.85\textwidth]{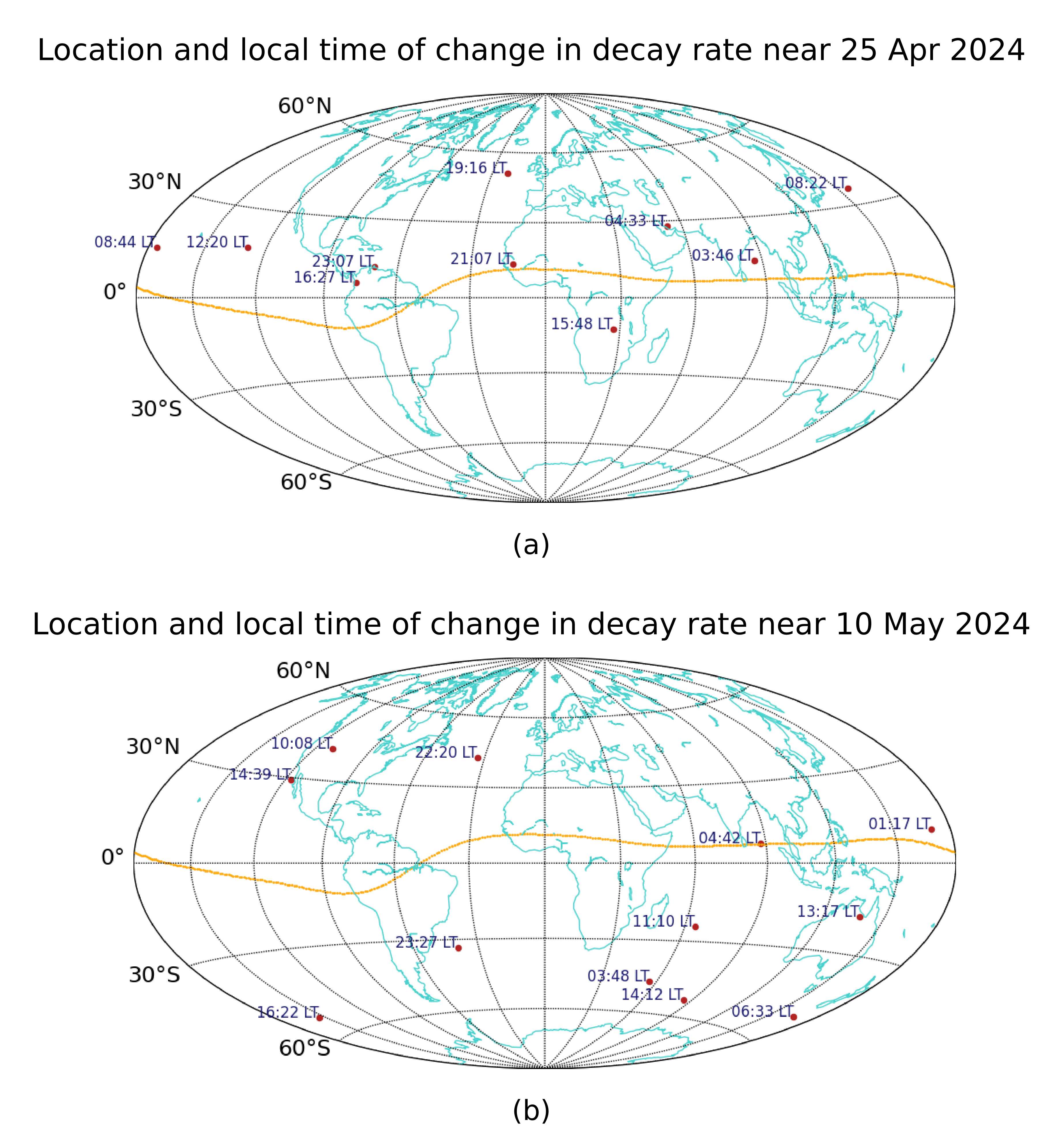}
	\caption{(a) Coordinates of change in the rate of decay near 25 April 2024 (b) coordinates of change in the rate of decay near 10 May 2024}
	\label{fig:Fig6b}
\end{figure}

Figure \ref{fig:DecayProfile} top panel depicts the orbital decay profiles of the 12 Starlink satellites along with Dst indices between 16 April and 15 May 2024. Following the geomagnetic storm on 10 May 2024, the satellites experienced sharp orbital decay due to increased thermospheric density. However, an unexpected increase in decay was observed around 25 April 2024, in 10 of the satellites at altitudes above 320 km. Two satellites, at initial altitudes of around 320 km, did not exhibit this pre-storm decay.  It was seen that this unexpected increase in decay coincided with a series of M class flares on the Sun. The NORAD (North American Aerospace Defense Command) IDs of these objects and the values of their rates of decay during distinct time intervals are listed in Table \ref{tab:Tab1}. Figures \ref{fig:Fig6b}a and \ref{fig:Fig6b}b show the positions of the satellites on a Hammer projection of Earth, when sudden changes in the rate of decay occurred near 25 April 2024 and 10 May 2024 respectively, based on TLE data.

A few interesting things can be noted from the top panel of Figure \ref{fig:DecayProfile}. Satellite 57649, initially at an altitude of approximately 400 km, experienced a decay rate of 0.7 km per day until around 26 April 2024. After this, the decay rate increased to about 8.1 km per day, and finally, there was a sharp rise to 45.3 km per day from 6 May 2024, leading to its reentry into the atmosphere. It must be noted that the second change in slope for this object occurred before the storm, unlike the other 11 satellites, where the second change in the rate of decay occurred after the storm's commencement.

Satellites 44951 and 44928, both starting at an altitude of 320 km on 16 April 2024, experienced constant decay rates of 2.2 km/day and 2 km/day, respectively. However, their decay rates sharply increased to 49.5 km/day and 57.4 km/day from 14 May 2024, leading to their reentry. The remaining objects exhibit a similar trend of sharp decay occurring up to 3 days after the main phase, except for object 48384, where the sharp decay aligns exactly with the main phase of the storm.

The sharp increase in altitude decay following the geomagnetic storm is expected; however, the rise in decay rates before the storm is unexpected. To investigate the enhanced orbital decays observed around 25 April 2024, we analyze EUV flux data from the EXIS/EUVS on board GOES. Figure \ref{fig:DecayProfile}, bottom panel, shows an EUV irradiance enhancement until 24 April 2024, which coincides with the increased decay rates of 10 satellites during this period, after which the EUV flux steadily decreases until 05 May. Solar EUV radiation plays a significant role in altering thermospheric neutral density. \cite{Briand2021} reported that EUV enhancements slow the thermosphere's recovery following solar flare–induced disturbances, based on CHAMP and GRACE accelerometer data. They found that these enhancements create a `density plateau', characterized by a sustained increase in neutral density of 30–70\% above normal levels that persists for several hours. Such plateaus can have a significant effect on satellite drag.

\begin{figure}[ht]
	\centering
	\includegraphics[width=0.85\textwidth]{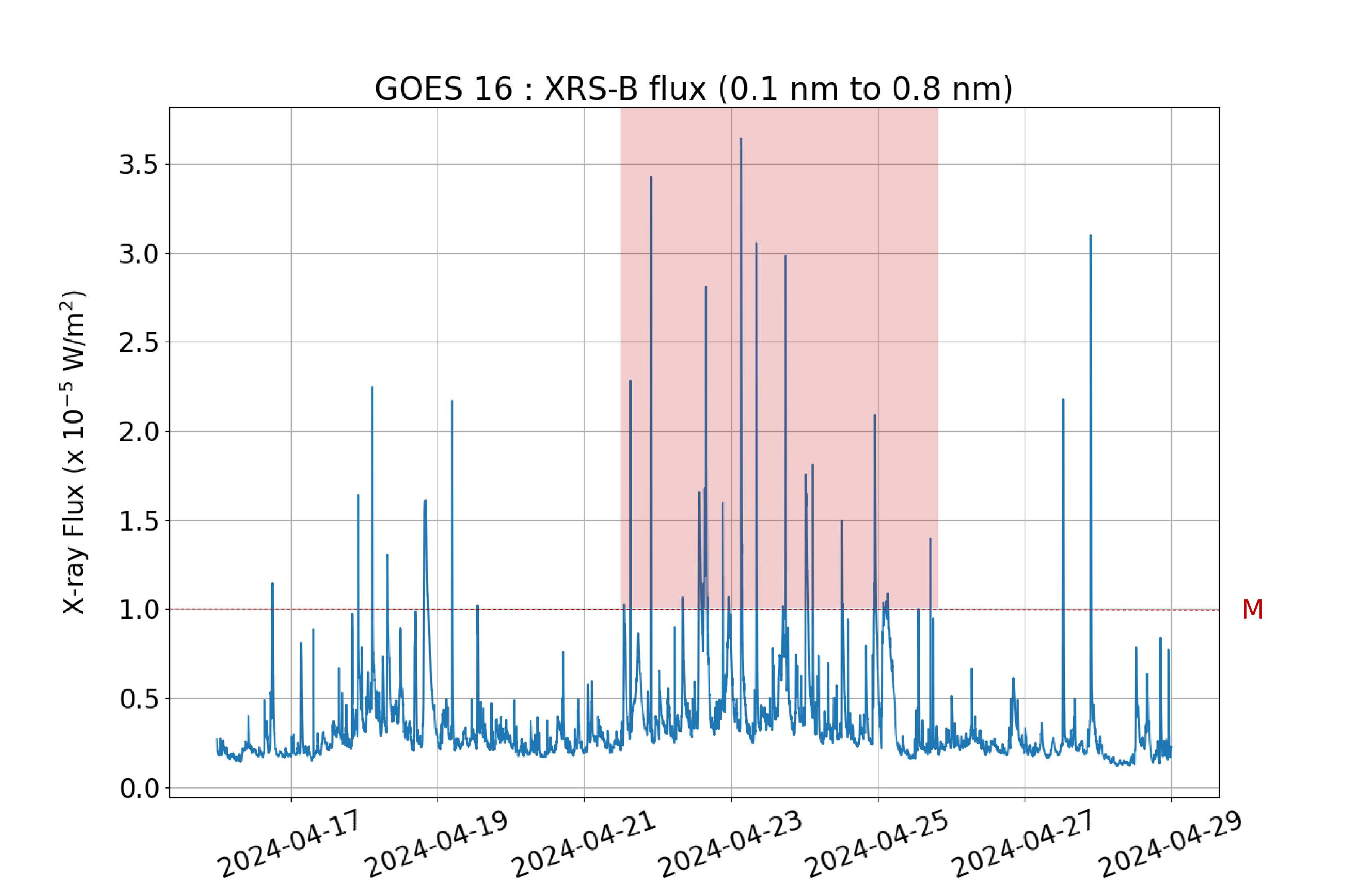}
	\caption{X-ray flux in the 0.1 nm - 0.8 nm band observed by the XRS instrument on GOES-16}
	\label{fig:GOESXRay}
\end{figure}

Additionally, between 22 April and 26 April 2024, a series of M-class flares occurred on the Sun, including an M3.5-class flare on 24 April 2024. Figure \ref{fig:GOESXRay} depicts the 1-minute averaged X-ray flux data from 16 April to 29 April 2024 in the 0.1 nm to 0.8 nm wavelength band, captured by the GOES XRS instrument. At LEO altitudes, thermospheric density can increase by up to 60\% in response to an X-class flare \citep{Qian2021}, leading to increased orbital decay of satellites.

\begin{figure}[ht]
	\centering
	\includegraphics[width=0.95\textwidth]{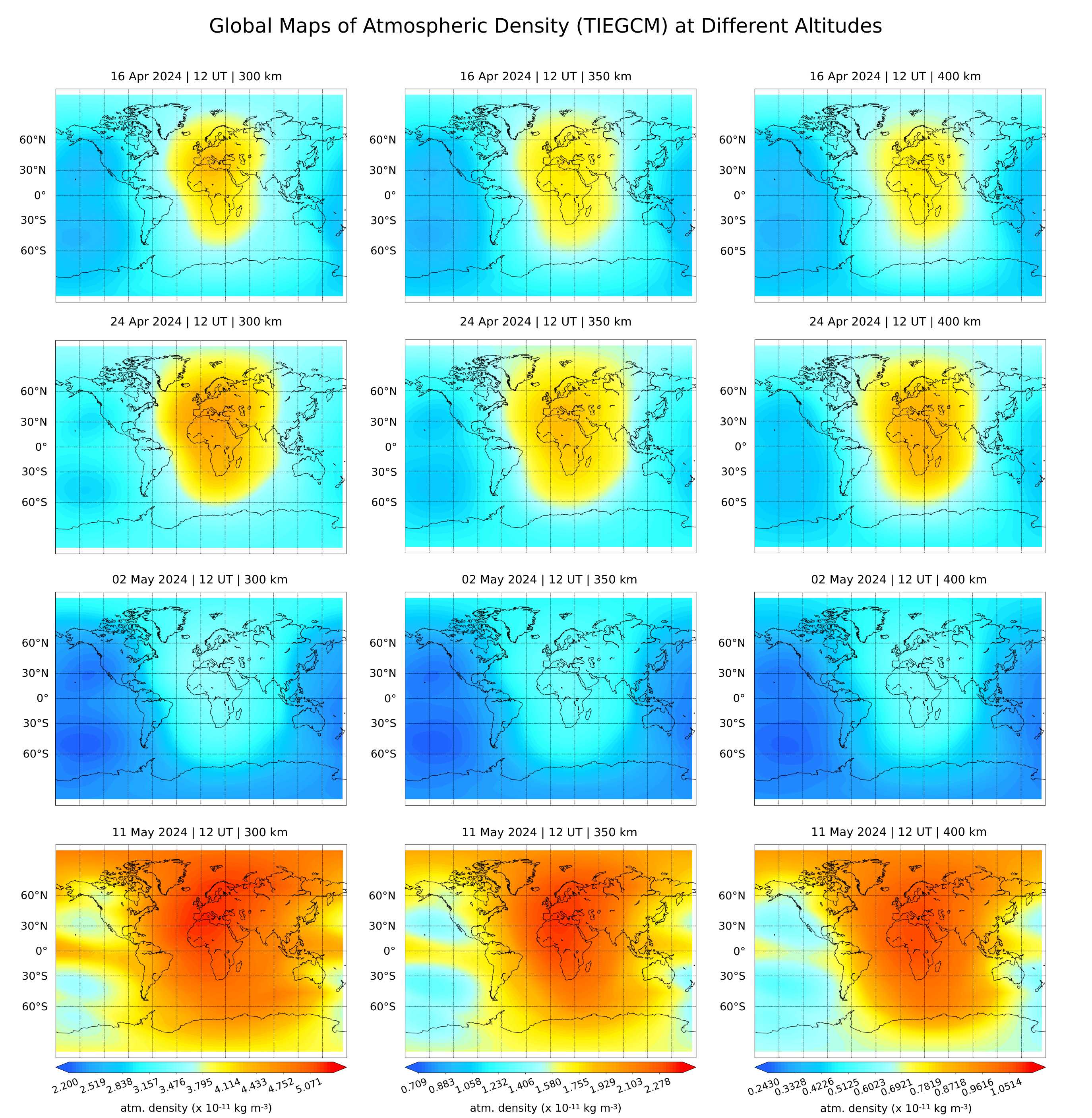}
	\caption{Global atmospheric density maps at altitudes of 300 km, 350 km, and 400 km for 16 April 2024 (pre-storm), 24 April 2024 (EUV enhancement with M-class flares), 02 May 2024 (pre-storm), and 11 May 2024 (storm main phase)}
	\label{fig:TIEGCM}
\end{figure}

Figure \ref{fig:TIEGCM} presents atmospheric density maps derived from the Thermosphere-Ionosphere Electrodynamics General Circulation Model (TIE-GCM), a first-principles, three-dimensional model that simulates the coupled thermosphere-ionosphere system by solving the momentum, energy, and continuity equations for neutral and ionized species on a staggered vertical grid extending from  around 97 km to 500 km. The model inputs include solar EUV radiation represented by the F10.7 index, particle precipitation obtained from the 3-hour Kp index, and high-latitude ionospheric electric fields at high latitudes specified by the Heelis and Weimer models \citep{Qian2013}. Densities between 16 April and 11 May 2024, obtained through the Community Coordinated Modeling Center (CCMC) TIEGCM model run, showed notable variations due to changes in solar activity. On 16 April, during quiet period, atmospheric density was moderate, with the peak region slightly shifted towards the Northern Hemisphere, as expected during the summer months. By 24 April, atmospheric density increased significantly, especially in the sunlit hemisphere, due to enhanced EUV flux and solar flares. This increase was most pronounced in the equatorial and mid-latitude regions. On 02 May, following the EUV and flare period, atmospheric density dropped sharply. Finally, on 11 May, during the main phase of the geomagnetic storm, atmospheric density saw a sharp increase, particularly at higher latitudes near the poles, due to intense energy input into the upper atmosphere, leading to substantial heating and expansion.

\begin{figure}[ht]
	\centering
	\includegraphics[width=0.85\textwidth]{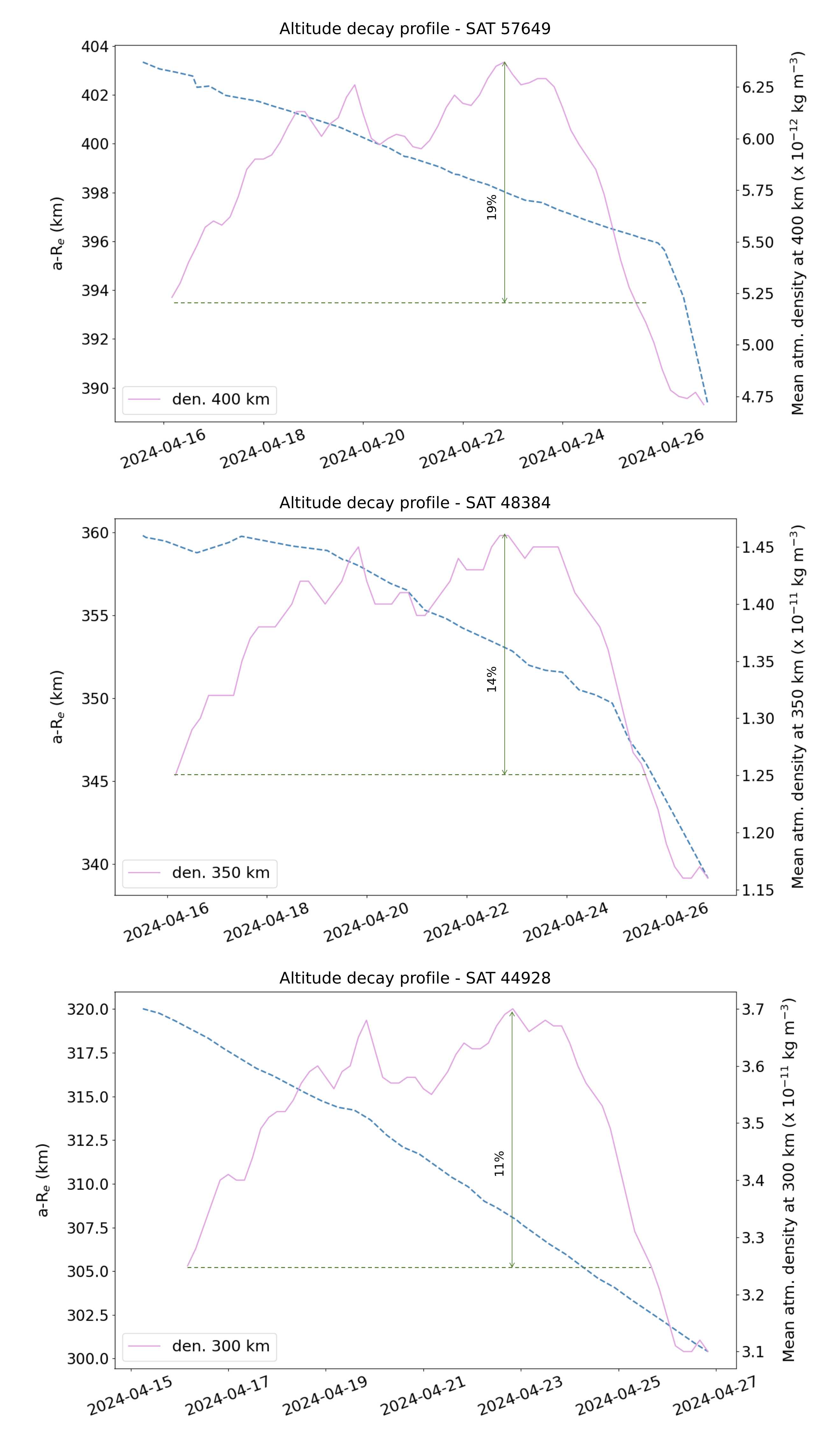}
	\caption{Decay profiles from 16 April 2024 to 27 April 2024 for satellites 57649 (400 km altitude regime),48384 (350 km altitude regime) and 44928 (300 km altitude regime) along with the corresponding global mean atmospheric densities derived from TIEGCM}
	\label{fig:densdecay}
\end{figure}

Figure \ref{fig:densdecay} shows the decay profiles of three sample satellites—57649 starting at 400 km, 48384 at 350 km, and 44928 at 300 km—along with the corresponding global mean atmospheric densities at those altitudes. Between 19 April and 24 April, density remains elevated across all altitudes before dropping sharply. The percentage increase in density from 16 April to the 23 April peak is smallest at 300 km (11\%), followed by 350 km (14\%), and highest at 400 km (19\%). After 24 April, decay rates increases to 5 km/day at 350 km and 8 km/day at 400 km, while the decay rate at 300 km remains steady at around 2 km/day.

\begin{figure}
	\centering
	\includegraphics[width=0.8\linewidth]{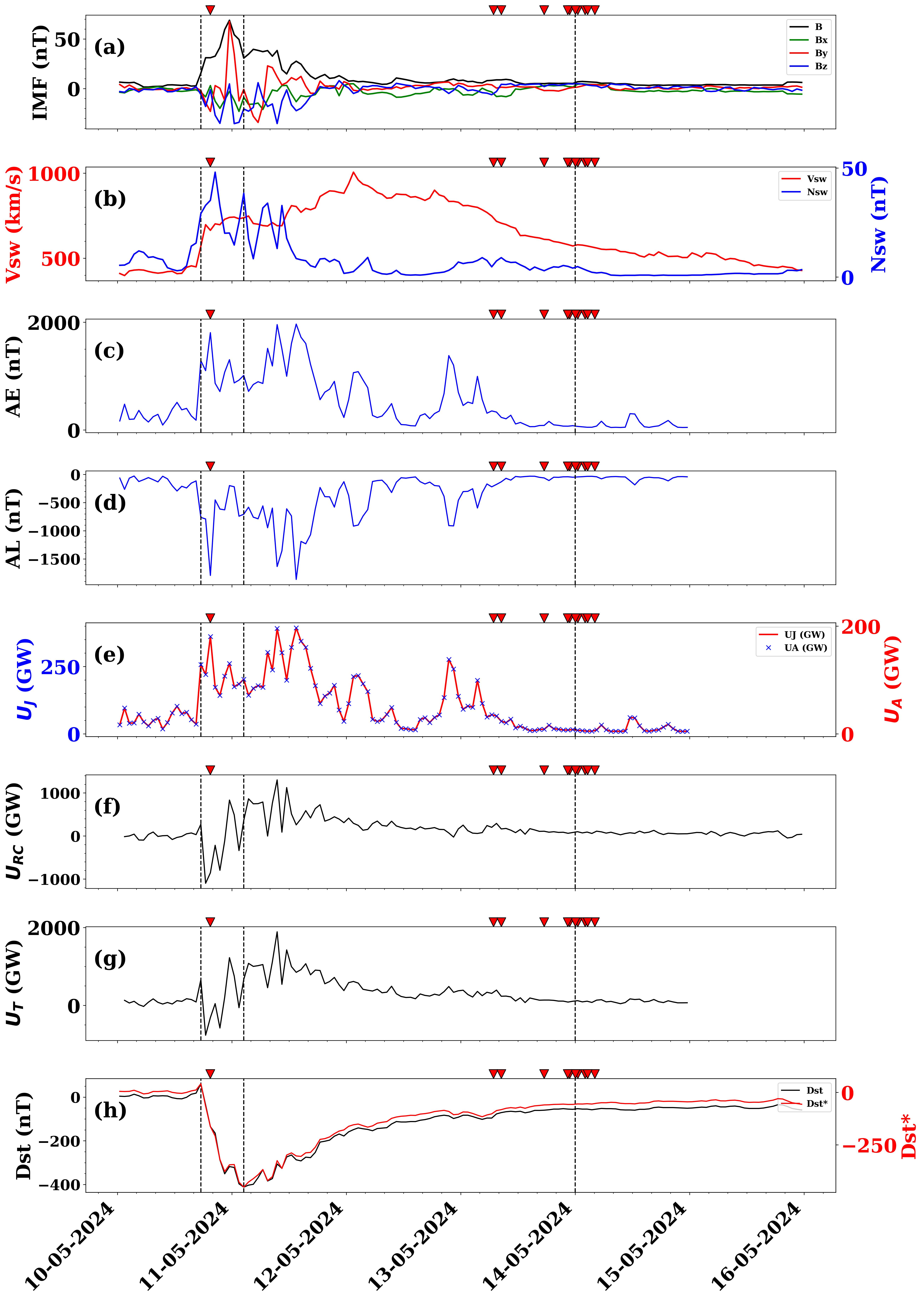}
	\caption{Various geophysical parameters associated with the geomagnetic storm. The three vertical dashed lines represent key phases of the storm: the leftmost line marks the onset of the storm and the beginning of the main phase, the second line indicates the minimum Dst index and the end of the main phase, and the rightmost line signifies the start of the recovery phase where Dst variations stabilize. Triangular symbols correspond to observations of decayed Starlink satellites, based on TLE data, reflecting the storm's impact on satellite trajectories.}
	\label{fig:energy_budget}
\end{figure}

Figure \ref{fig:energy_budget} represents the variations of hourly OMNI data \citep{OMNIdata}. The variations in the provisional disturbance storm time (Dst) index clearly show the main phase and recovery phase intervals, which are marked by dotted vertical lines. In panel `a', the magnitude of the interplanetary magnetic field (IMF) and its components Bx, By, and north-south component Bz are presented. The IMF magnitude sharply increased from 1.6 nT at 05:30 UT to a maximum of 68.9 nT at 23:30 UT on 10 May 2024. It remained significantly elevated throughout 11 May, not decreasing below 10 nT until the early hours of 12 May. Throughout most of this period, the Bz component remained primarily oriented southward (negative), reaching values as low as –35 nT at 21:00 UT on 10 May, and at 00:00 UT and 09:00 UT on 11 May. The persistently southward IMF component drove continuous magnetic reconnection at the dayside magnetopause, enhancing plasma convection within the inner magnetosphere and polar cap. From panel `b' of this figure, it can be seen that the solar wind speed increased continuously from 398 km/s on 10 May 2024, at 01:30 UT to a maximum value of 1006 km/s on 12 May 2024. When the Dst and the peak IMF field strength reached their minimum, the solar wind speed was 738 km/s on 11 May at 02:30 UT. The solar wind speed then slowly started decreasing and reached a minimum of 428 km/s on 15 May at 23:30 UT. The solar wind density started increasing until it reached a maximum value of 48.1 cm$^{-3}$ on 10 May 2024, at 20:30 UT.

The main phase which lasted for 9 hours is characterized by very strong intensity, as observed in the Dst index with a minimum of -461 nT. Panels `e', `f', and `g' represent the various energy estimates of the intense storm that occurred on 10 May 2024. The ring current injection rate U$_{RC}$,  joule heating of the magnetosphere U$_{J}$, and auroral precipitation U$_{A}$  are the major forms of energy dissipation in the magnetosphere.
The Dst index is influenced by magnetopause currents; therefore, in this study, we have corrected the Dst values using solar wind ram pressure to remove this contamination \citep{burton1975empirical,vichare2005some}.

\begin{equation}
	Dst^* =Dst-b \sqrt{P} + c
\end{equation}

where, P is solar wind dynamic pressure, and the coefficients
are set to b = 7.26 nT and c = 11 nT \citep{wang2003influence,o2000empirical}.
The total magnetospheric energy consumption rate can be determined by using quantitative estimation given by  \cite{akasofu1981energy}. The total energy contains a major contribution from three
components - U$_{J}$ ,U$_{A}$ and U$_{RC}$ \citep{akasofu1981energy, vichare2005some} which are defined by the following equations.

\begin{equation}
	U_{T} = U_{J}+ U_{A} + U_{RC} 
\end{equation}

\begin{equation}
	U_{J} = 2\times {10^8}{}AE
\end{equation}

\begin{equation}
	{U_{A}} = 1\times{10^8}AE
\end{equation}

\begin{equation}
	U_{RC} = -4 \times 10^{13} \left( \frac{d(Dst^{*})}{dt} + \frac{Dst^{*}}{\tau} \right)
\end{equation}

where $\tau$ is the decay time constant, set to 8 hours for intense geomagnetic storms \citep{yokoyama1997statistical}.

Panel `e' shows that the maximum value of U$_{J}$  and U$_{A}$ are  393 GW and 196 GW, respectively, on 11 May 2024 at 13:30 UT. Panel `f' shows a minimum value of  U$_{RC}$ - 1109 GW on 10 May 2024 at 18:30 UT and a maximum value of 1333 GW on 11 May at 09:30 UT. Panel `g' shows the minimum value of U$_{T}$ - 778 GW on 10 May at 18:30 UT and a maximum value of 1919 GW on 11 May at  09:30 UT. On 11 May 2024, at 02:30 UT when Dst reached its minimum value, the U$_{J}$, U$_{A}$, U$_{RC}$, U$_{T}$ were 202 GW, 101 GW, 357 GW, 661 GW respectively. The AE index was 1014 nT. The energy estimates show that a huge amount of energy was deposited into the magnetosphere-ionosphere system, which resulted in enhanced thermospheric heating during this intense storm. Note that the provisional nature of the geomagnetic indices used in this work may influence the estimation of energy input into the magnetosphere-ionosphere system, however, general observations are expected to remain the same.

\section{Discussion}

The observed orbital decay and subsequent reentries of 12 Starlink satellites during the geomagnetic storm on 10 May 2024 provide valuable insights into the effects of space weather on LEO satellites. While the post-storm increase in orbital decay was expected, the unexpected increase in decay rates for 10 satellites around 25 April 2024 required further analysis. A notable observation is that the two satellites that did not show a change in their orbital decay rates around 25 April were orbiting at altitudes below 300 km, while the other 10 satellites that did experience this sudden increase in decay rates were at altitudes of 350 km and 400 km. This decay could have been primarily driven by enhanced flare activity and EUV radiation on the Sun which caused density enhancements in the upper thermosphere. 

EUV radiation primarily affects the upper thermosphere by driving increased ionization and heating, leading to atmospheric expansion and density enhancements. \cite{Briand2021} used accelerometer data from GRACE and CHAMP satellites to measure density changes and showed that increase in EUV flux creates a `density plateau', characterized by a sustained increase in neutral density of 30–70\% above normal levels that persists for 6 to over 12 hours. CHAMP, which orbited at around 400 km experienced density increases of up to around 60\%. They demonstrated that the density perturbation produced and maintained by EUV enhancements is comparable to that observed during major solar flare events—but with a much longer recovery period. This means prolonged exposure to elevated drag conditions for LEO satellites. Therefore, underestimating these EUV-driven density increases may lead to substantial errors in orbital decay estimates. The temporal correlation between the EUV irradiance enhancement (Figure \ref{fig:DecayProfile}) and the observed increase in decay rates of the ten satellites suggests that solar EUV radiation played a direct role in thermospheric heating, subsequent density enhancement, and the resultant increase in drag forces. Additionally, Figure \ref{fig:DecayProfile} top panel shows that a moderate geomagnetic storm occurred on 19–20 April 2024, with a minimum Dst index of –117 nT recorded at 19 UT on 19 April.

In addition to the direct effects of EUV radiation, other ionospheric processes contribute to atmospheric density variations.  The equatorial anomaly refers to the anomalous latitudinal distribution observed in both the ionized and neutral components of the atmosphere. Specifically, the equatorial ionization anomaly (EIA) describes this irregular distribution in ions. The electron crest region exerts drag on the neutrals, leading to an increase in their density in the surrounding area \citep{Oigawa2021}, which causes the neutral anomaly. During periods of elevated solar flux, enhanced EUV radiation boosts ion production in the equatorial F region. This, together with intensified E × B drifts, strengthens the fountain effect - lifting plasma away from the equator and deepens the equatorial trough while enhancing the anomaly crests. Hence, the neutral anomaly also becomes more pronounced at higher solar flux levels \citep{Liu2007}. This trend is evident in Figure \ref{fig:DecayProfile}, which shows an enhancement in EUV flux over this time period. Figure \ref{fig:Fig6b}a depicts the locations and local time of the first change in the decay rates of the 10 satellites. Satellites 48599, 54001, 48430, 48579, and 48308 appear to have been affected by this phenomenon when passing through low latitudes (Figure \ref{fig:Fig6b}a). 

As shown in Figure \ref{fig:densdecay}, at lower altitudes (~300 km), satellites already experience significant atmospheric drag under normal conditions. Consequently, the additional increase in atmospheric density of around 11\% near 24 April had only a modest effect on their decay rates. In contrast, at higher altitudes (350–400 km), where the baseline atmospheric density is much lower, a slightly higher enhancement in density (14\% at 350 km and 19\% at 400 km) led to a proportionally larger increase in drag force, resulting in a marked acceleration of orbital decay rates following 24 April.

Joule heating, possibly caused by auroral electrojet currents flowing through the ionosphere, is a primary driver of thermospheric energy deposition. It is the second most significant energy sink following the ring current \citep{Baumjohann1984, Chun2002}.  Joule heating is most prominent at E-region altitudes over the poles, where Pedersen conductivity peaks \citep{Huang2012}.  Figure \ref{fig:energy_budget} depicts the interval centered around the geomagnetic storm showing incident energy in the ionosphere due to Joule heating (U$_{J}$) and auroral particle precipitation (U$_{A}$). The figure shows a significant energy input to the ionosphere during the storm. As a result, density perturbations at LEO altitudes are mainly driven by the upwelling of neutral particles and the expansion of the thermosphere due to heating pressure from below. This thermospheric expansion increases atmospheric density at LEO altitudes, resulting in higher drag forces on satellites, with the effect being most pronounced at mid to high latitudes \citep{Lu2016,Billett2024}.

The geomagnetic storm of 10 May 2024 was followed by substantial atmospheric disturbances. This included extreme auroral extensions observed as a result of the storm, with the auroral oval extending to below 30$^{\circ}$ magnetic latitude \citep{hayakawa2024}. The precipitating particles in the auroral region and intense heating due to auroral electrojets possibly contributed to ionospheric heating.  The resulting sudden and significant increase in thermospheric densities caused the rapid decay of satellites following the storm. However, there is a time lag of over 24 hours between the onset of the storm and the sudden change in the decay rate of 10 satellites. This heating of the atmosphere likely occurs, at least to some extent, through the movement of atmospheric waves and intricate north-south wind patterns, which require finite time to reach equatorial latitudes \citep{Oliveira2017, Fuller-Rowell1994}. \cite{Oliveira2019} reported a delay ranging from 12 to 72 hours for extreme to moderate geomagnetic storms, and the delay observed in this study falls within this range. For 10 of the satellites considered in this study, delays of up to 72 hours are seen in the occurrence of maximum decay. This order of delay is generally expected for a moderate storm (see \cite{Oliveira2019}), suggesting that the geo-effectiveness of the May 2024 storm was moderate, despite its G5 classification.

\section{Concluding Remarks}
The orbital decay and subsequent reentries of 12 Starlink satellites during and before the extreme geomagnetic storm of 10 May 2024 were examined. While the rapid post-storm increase in decay rate due to atmospheric disturbances was expected, the unexpected increase in decay rates before the storm, around 25 April 2024, proved interesting. It was observed that the EUV flux was significantly enhanced during this period, affecting the thermosphere through various processes. This indicated the important role of elevated space weather conditions before the storm, which resulted in increased satellite drag. These findings highlight the importance of understanding the complex interactions between solar activity, thermospheric dynamics, and satellite drag, particularly in the context of increased use of the LEO environment for space applications by the global community.

\bibliographystyle{jasr-model5-names}
\biboptions{authoryear}
\bibliography{refs}

\end{document}